\documentclass[twocolumn,showpacs,preprintnumbers,amsmath,amssymb]{revtex4}

\usepackage{graphicx}
\usepackage{dcolumn}
\usepackage{bm}

\begin{document}

\title{Topological instabilities of spherical vesicles}

\author{O. V. Manyuhina, A. Fasolino, P. C. M. Christianen and M. I. Katsnelson}%

\affiliation{Institute for Molecules and Materials, Radboud University Nijmegen, Heyendaalseweg 135, 6525 AJ Nijmegen, The Netherlands}

\date{\today}

\begin{abstract}
Within the framework of the Helfrich elastic theory of membranes and of differential geometry we study the relative stability of spherical vesicles and double bubbles. We find that not only temperature, but also magnetic fields can induce topological transformations between spherical vesicles and double bubbles and provide a phase diagram for the equilibrium shapes.
\end{abstract}


\maketitle

\begin{figure}[ht]
\centering
\raisebox{42mm}{(a)\hskip-2pt}\includegraphics[height=43mm]{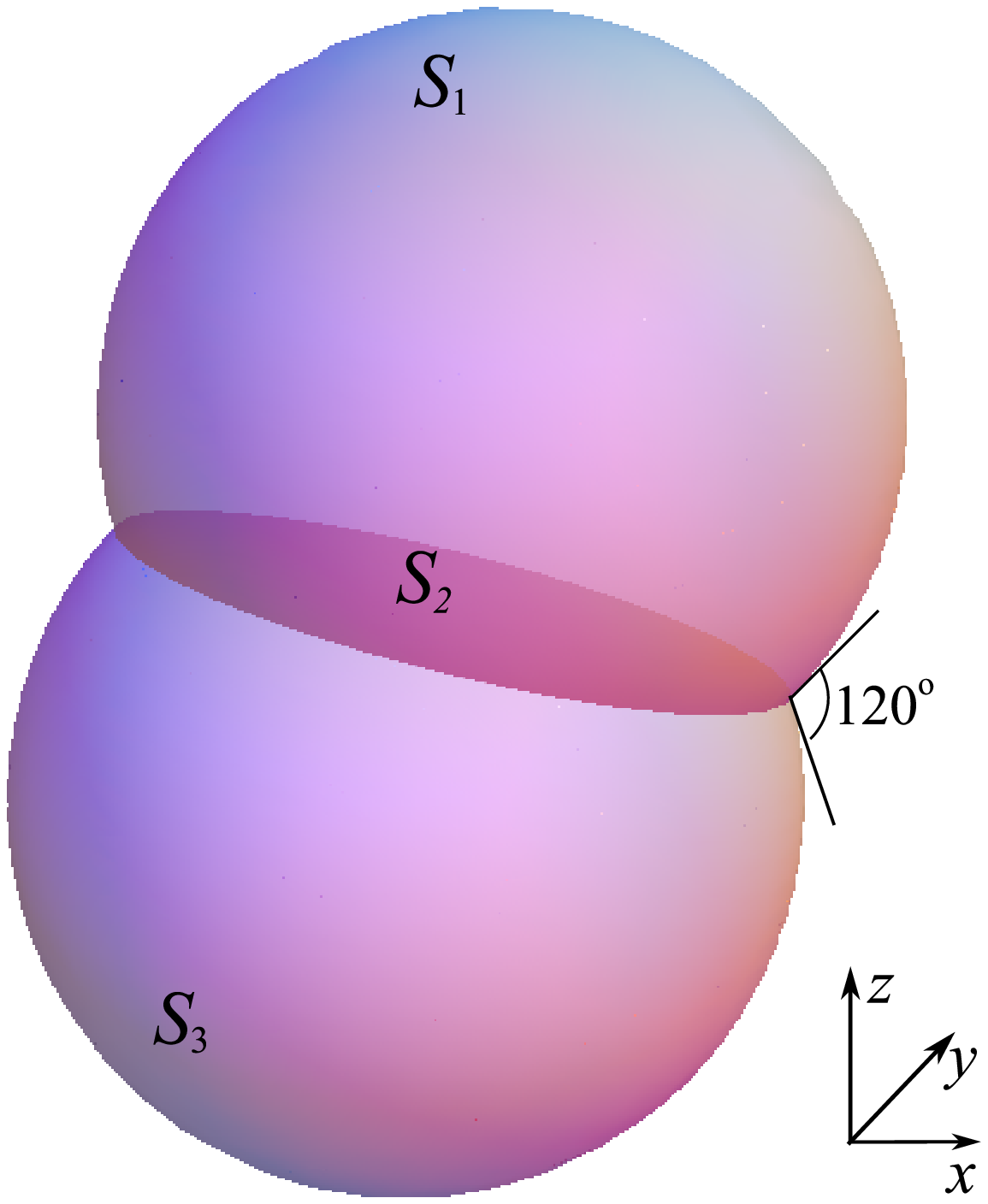} 
\hfil
\raisebox{42mm}{\hskip-6pt(b)\hskip3pt}\includegraphics[height=43mm]{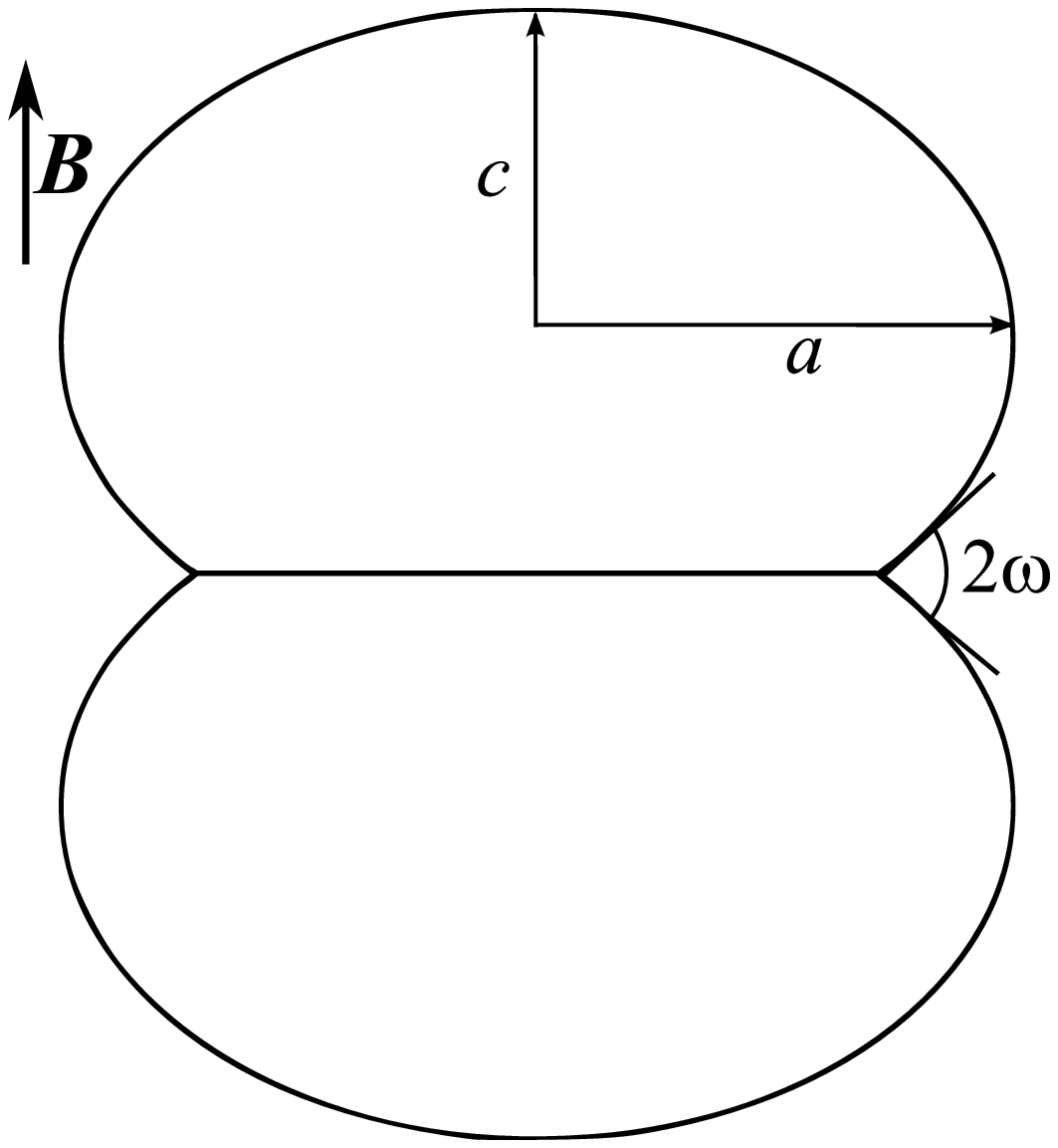}
\caption{(color online) (a) The double bubble is the unique surface of least area that encloses two equal volumes; it is composed of two identical spherical caps $S_1$ and $S_3$ separated by  a single membrane $S_2$ that is flat for equal volumes; the three pieces meet along a circle at $120^\circ$ angle~\cite{Hass}. (b) Cross-section of a magnetically deformed double bubble with semi-axes $a>c$; $2\omega$ is the angle between two tangent planes at the meeting circle.} \label{fig1}
\end{figure}

Self-assembled vesicles were experimentally observed to form a variety of shapes~\cite{Seifert}. However most equilibrium vesicles are spherical. This is not surprising, because spheres are known to minimize not only the surface energy but also the elastic energy of vesicles~\cite{Zhang}. In mathematics it has been proven that spheres have the least area to enclose and separate a given volume. The analogous problem of identifying the surface with least area enclosing and separating more than one volume was first considered by the Belgian physicist J.~Plateau in the XIXth century~\cite{Plateau}. Surprisingly, only recently it has been proven that a double bubble, a figure composed of two spherical caps separated by a flat disk (Fig.~\ref{fig1}a), is the solution to the two equal volume isoperimetric problem in Euclidean space~\cite{Hass}. A general proof for double bubble enclosing two different volumes separated by a non-flat membrane was given later~\cite{Morgan}.

The connected spherical bubbles  often observed in soap are certainly a realization of this `minimum property' of double bubbles since surface tension is the dominant driving force. In this paper we explore theoretically whether double bubbles can represent the equilibrium shape of vesicles when not only the surface tension, but also other energy terms like elastic and magnetic energies are relevant. To the best of our knowledge, a double bubble was never considered as a candidate to describe the shape of vesicles. The presence of the membrane $S_2$ (see Fig.~\ref{fig1}a) distinguishes double bubbles from dumbbells or pear-shaped vesicles that were extensively studied in phase separated systems~\cite{Lipowsky}. Two adhering vesicles may look similar to a double bubble, but they have different topology, because the membrane $S_2$ is composed of two interacting membranes~\cite{Svetina}. Hence, when vesicles adhere there is no change in topology and the Euler characteristic~\cite{Carmo} remains unchanged as for two single vesicles ($\chi=4$), whereas a double bubble has $\chi=3$~\cite{chi}, that is neither the topology of two single vesicles nor the one of a larger sphere ($\chi=2$). One may therefore  refer to a double bubble as an intermediate shape in the topological transformations between one and two spheres. Such transformations might also be relevant to describe the first stage of cell division.

To study the instabilities of spherical vesicles towards double bubbles we start from the well known model of elastic free energy proposed by Helfrich~\cite{Helfrich}. This model was successful in explaining the shapes of red blood cells and in predicting some new non-trivial shapes of vesicles~\cite{Seifert,Zhang}. However, one cannot apply without restrictions a continuum elastic theory to the shapes with discontinuities, since it would yield infinite energy. The discontinuities can be taken into account by introducing a line tension with an effective angular dependence, as was proposed in~\cite{BenAmar}. Similar arguments can be applied to a double bubble, which has a circular rim with a singularity, where three smooth surfaces meet (see Fig.~\ref{fig1}a). Therefore, to take into account the rim singularity in the shape of a double bubble, we propose a new phenomenological term derived from the Gauss--Bonnet theorem, and added to the Helfrich elastic free energy. Although other forms of the rim energy might be plausible, here we investigate the consequences of our ansatz. Assuming that intrinsic parameters of the system entering the phenomenological Helfrich free energy, like elastic moduli and spontaneous curvature, are temperature and pressure dependent~\cite{Shearman}, we propose a phase diagram for the studied shapes. It is intuitively clear that shapes similar to a double bubble may occur whenever  flat membranes become favorable, for instance in presence of high magnetic fields that tend to orient diamagnetic molecules. For this case, experimental evidence of deformation of self-assembled spherical vesicles to oblate spheroid  was reported in~\cite{Shklyarevskiy,2007}. Here, we find that a magnetic field can also influence the relative stability of shapes and can make two spheroidal vesicles less stable than a deformed double bubble or one larger spheroid. The latter transformation was experimentally observed  for liposomes~\cite{Ozeki} in high magnetic fields. 




In Fig.~\ref{fig1}a we show the double bubble as defined  mathematically. The double bubble is composed of two spherical caps $S_1$ and $S_3$ separated by a disk $S_2$, meeting along a common circle (rim)  at an  angle of $120^\circ$~\cite{Hass}. To consider magnetic deformations, we study the modified geometry shown in Fig.~\ref{fig1}b. Firstly, we assume that, during the  deformation, the spherical caps $S_1$ and $S_3$ (with $S_1=S_3$) change to oblate spheroids with semiaxes $c$ and $a$ parallel and perpendicular to the magnetic field $B$ respectively. Secondly, we allow the angle between the three surfaces to differ from $120^\circ$ by introducing the parameter $\omega$ (see Fig.~\ref{fig1}b), related to the radius of the membrane as 
\begin{equation}\label{eq:Rm}
R_{\rm m} = \frac{a^2\sin\omega}{\sqrt{a^2\sin^2\omega +c^2\cos^2\omega}}.
\end{equation}
In this paper we extend the concept of double bubble from the unique area minimizing surface with $c=a$ and $\omega=\pi/3$ to the set of shapes with $c\in(0,a]$, $\omega\in(0,\pi/2)$.

The equilibrium shapes of fluid membranes are usually studied in terms of the Helfrich model of bending energy~\cite{Helfrich,Safran}:
\begin{equation}\label{eq:Fel}
{\cal F_{\rm el}}=\int \!dS\, \{2k (H-H_0)^2+ \bar k K\},
\end{equation}
where $H$ and $K$ are the mean and Gaussian curvatures respectively, and $H_0$ is the spontaneous mean  curvature. The first term, proportional to the bending rigidity $k$, describes the local deviation from equilibrium curvature $H_0$, while the second term with Gaussian rigidity $\bar k$ influences only the topology and thus is often neglected. For the same reason, $\bar k$ is also not measurable if no topological transformations occur. Here we cannot omit this term, because we examine different topologies. It was shown theoretically and confirmed experimentally that $-1<\bar k/k<0$ for monolayer vesicles, while for bilayer vesicles $\bar k$ can be positive~\cite{Shearman,Zasad}. The values of $H_0$, $k$ and $\bar k$ depend on pressure and temperature, and therefore may be considered as parameters and not as intrinsic properties of the system. 
We will consider the case with $0.7<H_0R_0<1.3$, $-1\leq\bar k/k\leq1$, where $R_0$ is the radius of the spherical vesicle, and $k=4\cdot10^{-20} J\approx10~k_{\rm B}T$, the latter being a typical value for molecular vesicles. Moreover, we assume that self-assembled molecules keep a constant density, implying the condition of constant surface of the vesicles $S=8\pi R_0^2$.

The form of the phenomenological free energy, given by Eq.~(\ref{eq:Fel}), is valid only for smooth surfaces like spheroids or the individual surfaces $S_1, S_2$ and $S_3$. In order to introduce the energy of the rim phenomenologically, we make use of the Gauss--Bonnet theorem $\int dS\,K=2\pi\chi$~\cite{Carmo}, where $\chi=3$ for the double bubble~\cite{chi}. By integrating explicitly the Gaussian curvature $K$ over the three smooth pieces of the double bubble, namely the spheroidal caps $\{S_1,S_3\}$ and the flat membrane $S_2$ ($K=0$), we find the following expression 
\begin{equation}\label{eq:K}
\sum_{i=1}^3\int_{S_i} dS\,K  = 4\pi(1+\cos \omega),
\end{equation}
which depends only on the angle $\omega$~(see Fig.~\ref{fig1}b). Since $\int dS\,K=6\pi$, the difference with respect to Eq.~(\ref{eq:K}) represents the contribution of the curvature integral over the rim. This integral accounts for the discontinuity of the shape and yields the angle deficit of the rim as $2\pi\chi-\sum_{i=1}^3\int_{S_i} dS\,K=2\pi(1-2\cos\omega)$. We then assume that  the phenomenological free energy of a circular rim is proportional to the square of the angle deficit $(1-2\cos\omega)$ multiplied by the rim length $2\pi R_{\rm m}$
\begin{equation}\label{eq:Frim}
{\cal F}_{\rm rim}=\gamma(1-2\cos\omega)^2 2\pi R_{\rm m},
\end{equation}
where the coefficient of proportionality $\gamma$ represents the line tension. The value of $\gamma$ can be estimated by considering the formation of vesicles due to edge effects and comparing the free energy of the two limiting cases, namely a disc with $F_{\rm disc}=2\pi\gamma R_{\rm d}=4\pi\gamma R_0$   and a sphere with $F_{\rm sphere} = 4\pi 2 k$, yielding $\gamma = 2 k/R_0$~\cite{Antonietti}. Our conjecture for the form of the rim free energy ${\cal F}_{\rm rim}$ is similar to the line tension with an effective angular dependence proposed in~\cite{BenAmar} to justify the occurrence of slope discontinuities between different membrane domains. Also for crystals, it has been demonstrated mathematically in~\cite{Taylor}, that the cusps that are often present at crystal surfaces can be intrinsic property of equilibrium shapes and not necessarily due to defects.


\begin{figure*}
\centering
\raisebox{57mm}{(a)}\includegraphics[clip=true,height=60mm]{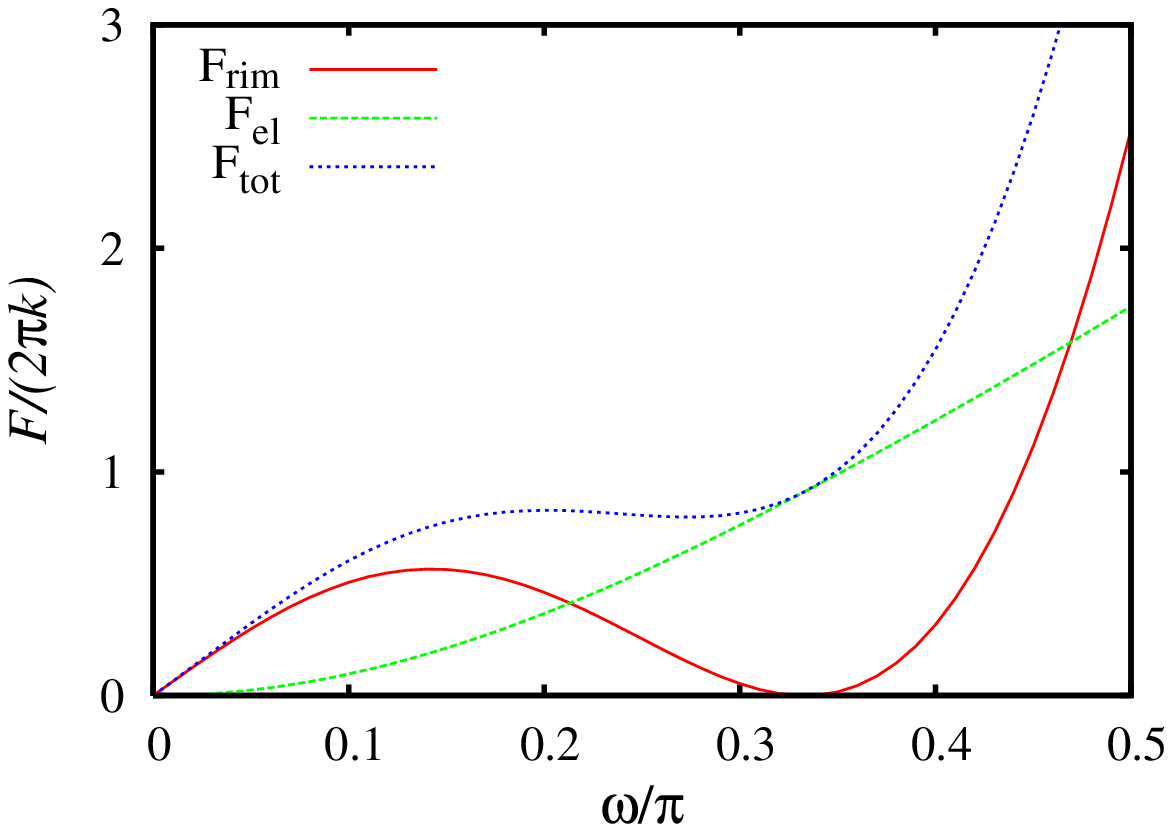}
\hfil
\raisebox{57mm}{(b)}\includegraphics[clip=true,height=60mm]{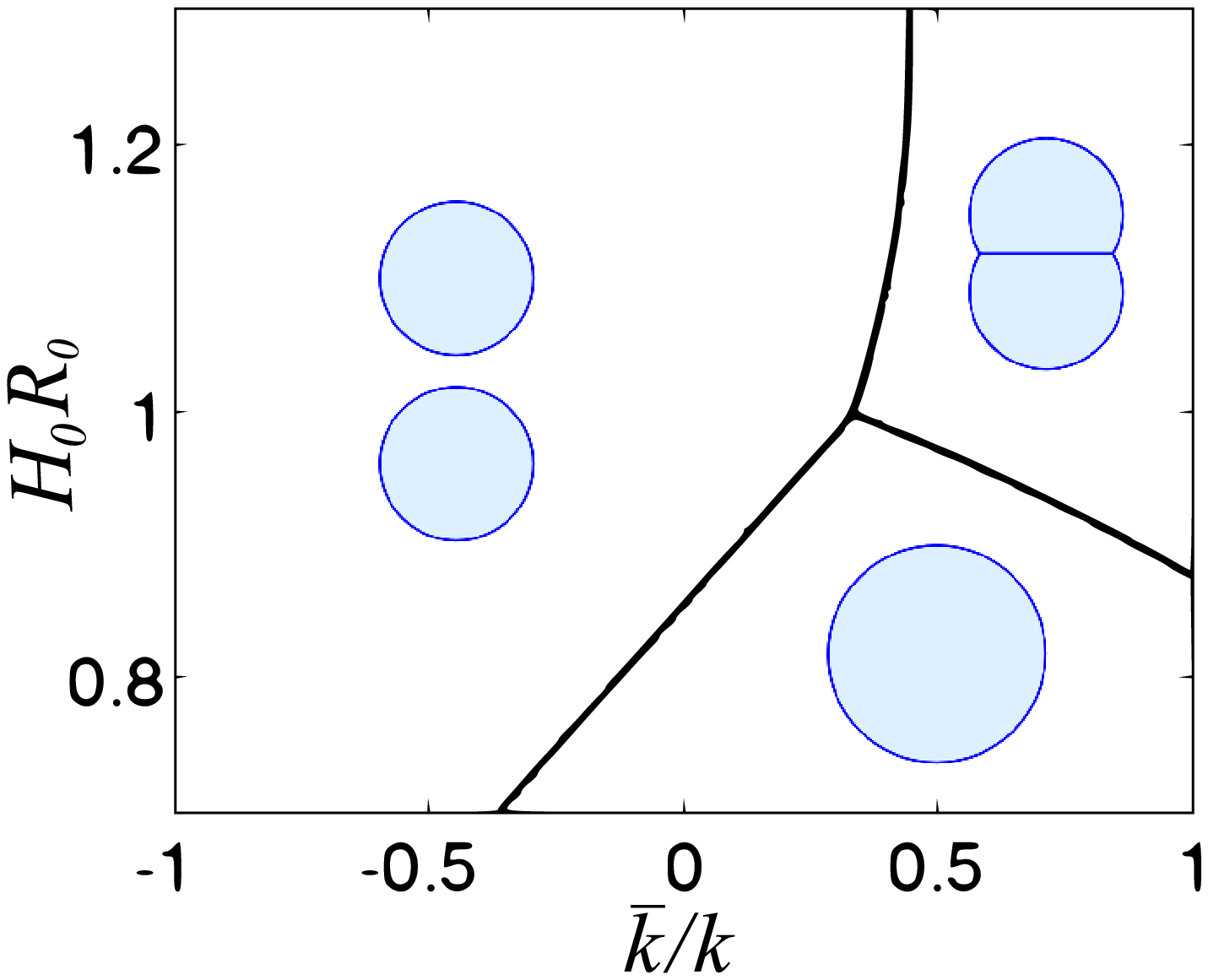}
\caption{(color online) (a) The free energy terms ${\cal F}_{\rm rim}$, ${\cal F}_{\rm el}$ and ${\cal F}_{\rm tot}$ of double bubble with $c=a$ plotted versus the angle $\omega$. We choose $\bar k=0$ and $H_0R_0=1$ for the elastic part for illustration. (b) Phase diagram showing regions where two spheres with radius $R_0$, a standard double bubble with $c=a$ and $\omega=\pi/3$, and one single vesicle with radius $\sqrt{2}R_0$ have the lowest free energy. The parameters $H_0$, $ k$ and $\bar k$ depend on temperature and pressure~\cite{Shearman}.}
\label{fig2}
\end{figure*}

\begin{figure}[ht]
\centering
\includegraphics[clip=true,width=0.6\linewidth]{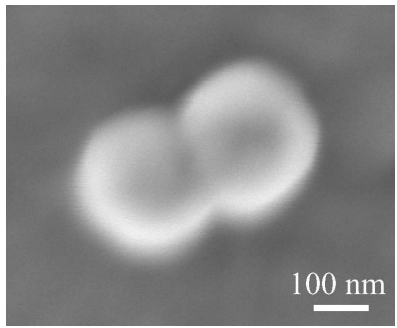}
\caption{Scanning electron microscopy (SEM) image of 6T vesicles on a metallic surface found after evaporation of a solvent.}
\label{fig3}
\end{figure}

We begin by considering the symmetric case, namely a double bubble composed of two spherical caps with\break ${c=a=R=R_0\sqrt{8/(4(1+\cos\omega)+\sin^2\omega)}}$, which yields $R|_{\omega=0} =R_0$. The result for the elastic energy and the rim energy is ${\cal F}_{\rm el}=4\pi  k(1+\cos\omega)[(1-RH_0)^2+\bar k/(2 k)] + 2\pi  k\sin^2\omega (RH_0)^2$ and ${\cal F}_{\rm rim}=4\pi  k\sin\omega(1-2\cos\omega)^2  R/{R_0}$. 
The $\omega$-dependence of these terms is shown in Fig.~\ref{fig2}a. The rim energy vanishes at $\omega=0$ when two vesicles just touch and at $\omega=\pi/3$ when the angle between three surfaces is $2\omega=2\pi/3$ (see Fig.~\ref{fig1}) that coincides with the Plateau's rule for soap films. However, contrary to soap bubbles, the elastic energy grows with $\omega$ so that the total free energy ${\cal F}_{\rm tot}={\cal F}_{\rm el}+{\cal F}_{\rm rim}$ does not necessarily have a minimum in the vicinity of $\omega=\pi/3$, depending on  the relative contribution from ${\cal F}_{\rm el}$. These results show that with this approach we can obtain the correct limiting cases. Figure~\ref{fig2}b presents the phase diagram for the relative stability of spherical vesicles and double bubbles in the plane of the parameters $(\bar k/ k)$--$(H_0R_0)$. Thermal fluctuations can be taken into account via the renormalization of the bending and Gaussian rigidity given by~\cite{David}\break $ k_{\rm eff}= k - (3k_{\rm B}T/8\pi)\log({q^2_{\rm max}}/{q^2_{\rm min}})$ and $\bar k_{\rm eff}=\bar k + \break(5k_{\rm B}T/12\pi)\log({q^2_{\rm max}}/{q^2_{\rm min}})$. Therefore, an increase in temperature might be enough to induce shape transformation. In Fig.~\ref{fig3} we show a SEM image of a structure formed by  self-assembled  bolaamphiphilic sexithiophene (6T) molecules that usually form hollow spherical vesicles~\cite{Shklyarevskiy}. The angle between two spherical parts in this image suggests the possibility that this structure is a double bubble and not two adhering vesicles. It would be interesting to apply other experimental techniques, such as TEM, to investigate this issue.




Now consider the effect of magnetic field, which tends to align diamagnetic molecules and usually leads to deformation of spherical vesicles towards superspheroid like the one measured and explained in~\cite{Shklyarevskiy,2007}. Here, we consider the possibility that magnetic fields could affect the topology as well. We minimize the total free energy 
\begin{equation}\label{eq:Ftot}
{\cal F}_{\rm tot} = {\cal F}_{\rm el}+{\cal F}_{\rm rim} + {\cal F}_{\rm mag}, 
\end{equation}
with respect to the deformation $c/a$ for a given value of~$\omega$. The elastic energy ${\cal F}_{\rm el}$ and the rim energy ${\cal F}_{\rm rim}$ are given by Eqs.~(\ref{eq:Fel}) and~(\ref{eq:Frim}), respectively, and the magnetic energy, assuming that the membrane $S_2$ is perpendicular to the direction of the magnetic field, is
\begin{equation}\label{eq:Fmag}
{\cal F}_{\rm mag} =-\frac{\Delta\chi D B^2}{2\mu_{0}}\bigg(\int \!dS\, n_z^2 +\pi R_{\rm m}^2\bigg),
\end{equation}
where  $\Delta \chi = \chi_\parallel -\chi_\perp$ is the difference of magnetic susceptibility along the long and short axes of the molecule and $D$ is the length of the molecule and the last term is the contribution of the flat disk with radius $R_{\rm m}$ (Eq.~(\ref{eq:Rm})). Whenever the reduction in magnetic energy ${\cal F}_{\rm mag}$ due to the presence of the membrane compensates the elastic energy ${\cal F}_{\rm el}$ and the energy of the rim ${\cal F}_{\rm rim}$, the formation of a double bubble is favored.


\begin{figure*}[ht]
\centering
\raisebox{57mm}{(a)}\includegraphics[clip=true,height=60mm]{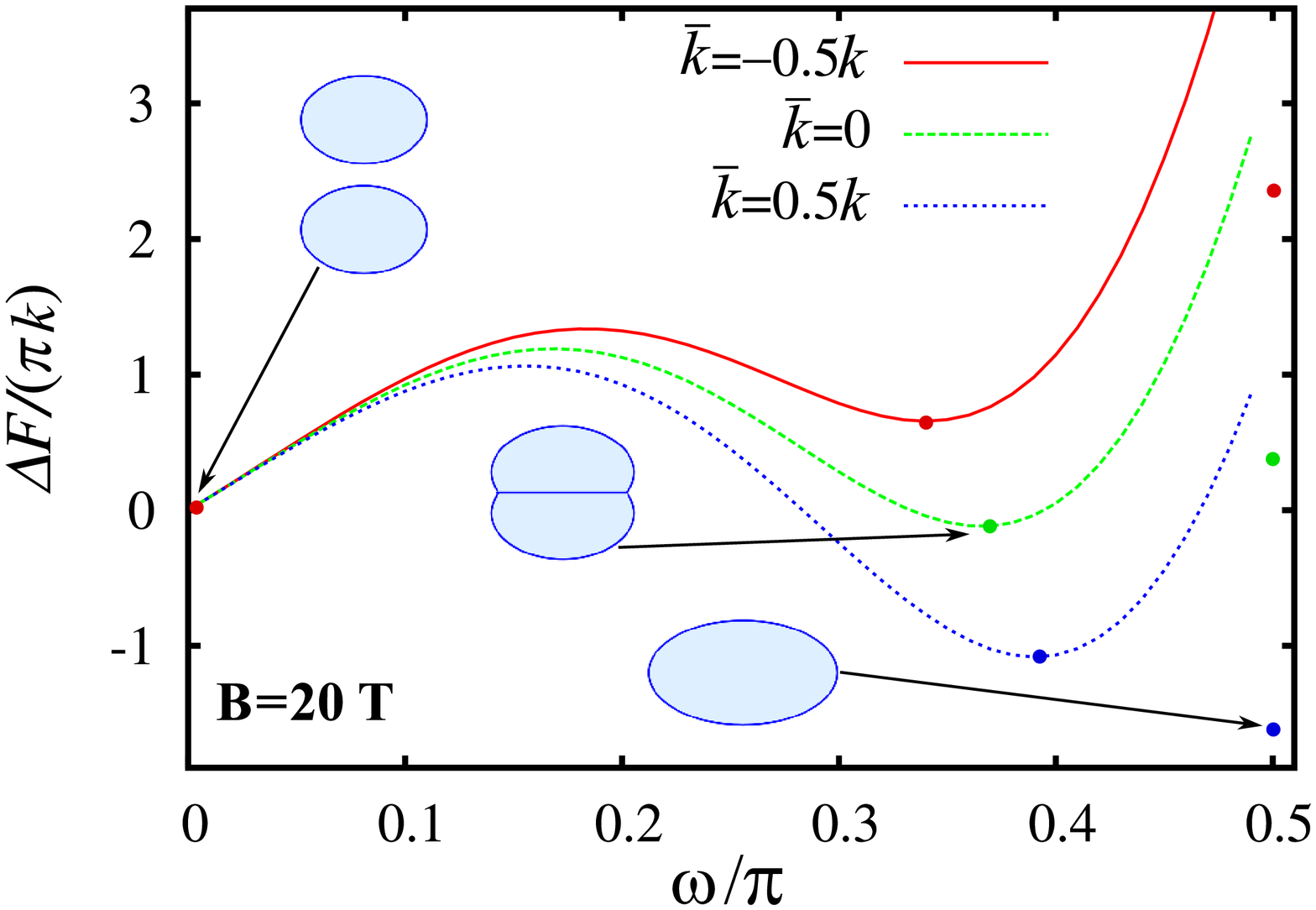}
\hfil
\raisebox{57mm}{(b)}\includegraphics[clip=true,height=60mm]{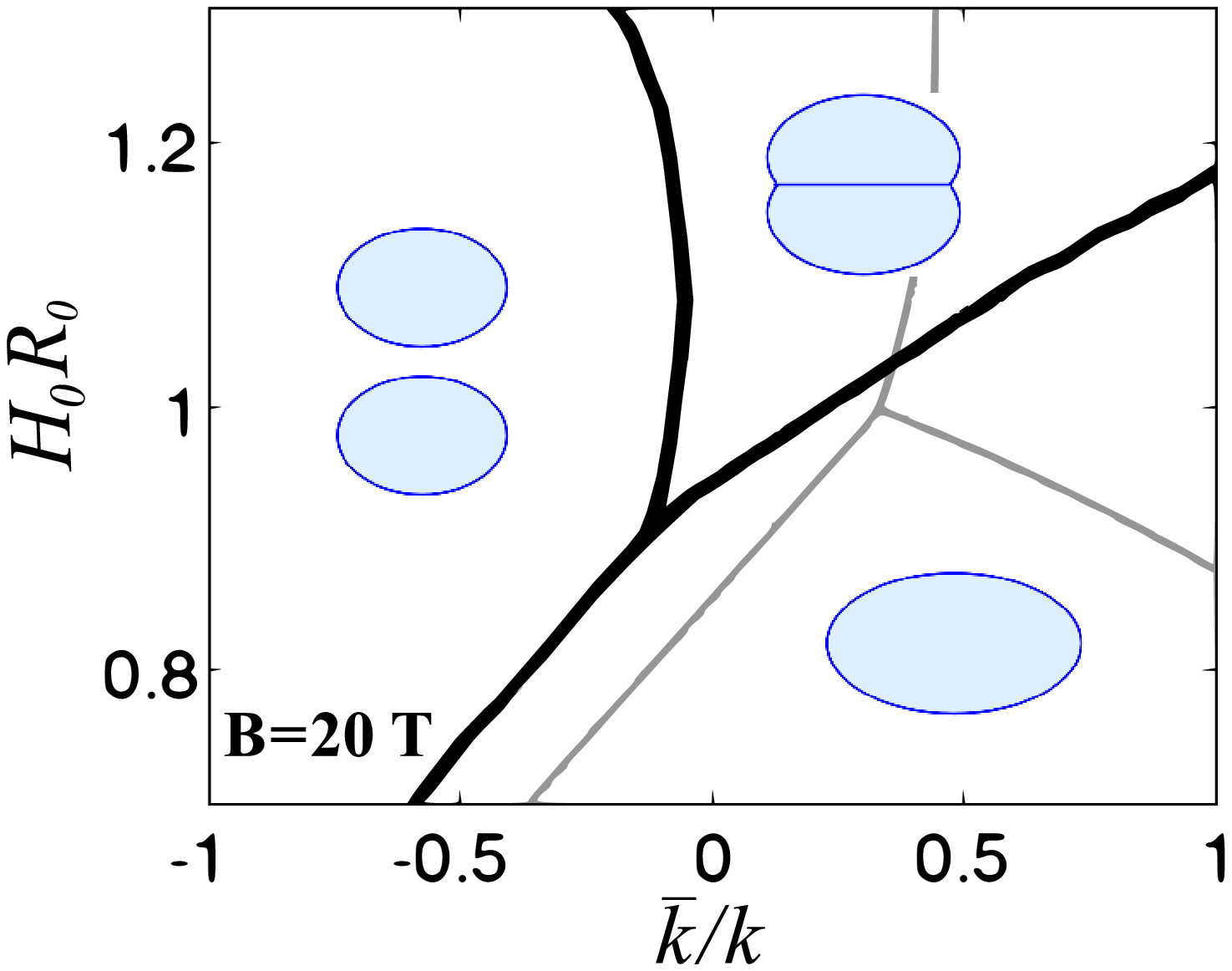}
\caption{(color online) (a) Variation of the minimum of the total free energy $\Delta{\cal F}=\min_{\omega={\rm const}}{\cal F}_{\rm tot}(c/a)-\min_{\omega=0}{\cal F}_{\rm tot}(c/a)$ for magnetic field $B=20$~T.  The thick dots on each curve indicate the local minima separated by barriers. For some cases we show the corresponding equilibrium shapes. For this plot the parameters are $H_0R_0=1$, $R_0=100$~nm, $D=5$~nm, $\Delta\chi =10^{-5}$. (b).~Phase diagrams enclosing three regions with topologically different shapes: thick black lines ($B=20$~T) and grey thin lines ($B=0$~T same as Fig.~\ref{fig2}b). Shapes with different topologies illustrate schematically the regions of the lowest free energy.}
\label{fig4}
\end{figure*}

We show first the free energy, minimized over the deformation $c/a$, for all possible values of $\omega$ in a magnetic field chosen as $B=20$~T (see Fig.~\ref{fig4}a). 
We distinguish three deep minima, separated by high barriers, corresponding to the three equilibrium shapes. By varying the  value of Gaussian rigidity we find as the ground state: two spheroids for $\bar k=-0.5 k$, a deformed double bubble with $\omega\simeq0.37\pi$ for $\bar k=0$, and a single spheroid for $\bar k=0.5 k$ with comparable values of deformation $c/a$. Nevertheless, the presence of three separated minima means that the experimentally observed shapes would depend on the kinetics of the system and that under certain conditions all three shapes could exist simultaneously. In Fig.~\ref{fig4}b we illustrate the phase behaviour of the studied shapes at $B=20$~T compared to the one at $B=0$~T in Fig.~\ref{fig2}b. As we expected, in presence of a magnetic field, the area in the phase diagram of deformed double bubbles and single spheroids increases significantly, which leads to a larger probability of finding these shapes for the fluctuating spherical vesicles. These results also show that the change in topology from sphere to double bubble can be induced not only by temperature but also by magnetic fields.


In conclusion, we found that the double bubble, the surface of smallest area enclosing two equal volumes, also minimizes, under certain conditions, the free energy of self-assembled elastic vesicles. We have explicitly shown that magnetic fields can be used to alter not only the shape and the size, but also the topology of two diamagnetic vesicles. Considering the possibility of other topologies is useful when analyzing experimental results and establishing whether self-assemblies are in equilibrium or not. The calculations presented in this paper may be thought as a first step to consider the formation of foam during self-assembly, like the one observed in~\cite{F}. We hope that our work will provide a motivation for new experiments.

It is a pleasure to acknowledge helpful discussions with Michael Mueger, Philippe Nozi\`eres and Efim Kats. 


\begin{thebibliography}{99}
\bibitem{chi}
The Euler characteristic of a polyhedron is defined as $\chi=F-E+V$, where $F$ is the number of faces, $E$ is the number of edges and $V$ is the number of vertices of the polyhedron. Since the double bubble is homeomorphic to two tetrahedrons with one common face, yielding $F=7$, $E=9$ and $V=5$, its  Euler characteristic is $\chi=3$.
\end{thebibliography}
\end{document}